\documentstyle[aps,prb,preprint]{revtex}
\begin{document}
\title{Ultrafast photoinduced reflectivity transients in (Nd$_{0.5}$Sr$_{0.5}$)MnO$%
_{3}$}
\author{T. Mertelj and D. Mihailovic}
\address{Jozef Stefan Institute, Jamova 39, 1000 Ljubljana, Slovenia\\
and}
\address{University of Ljubljana, Faculty of Mathematics and Physics,\\
Jadranska~19, 1000 Ljubljana, Slovenia}
\author{Z. Jagli\v{c}i\v{c}}
\address{Institute for Mathematics, Physics and Mechanics, Jadranska~19, 1000\\
Ljubljana, Slovenia}
\author{A. A. Bosak, O. Yu. Gorbenko and A. R. Kaul}
\address{Department of Chemistry, Moscow State University, Moscow 119899, Russia}
\maketitle

\begin{abstract}
The temperature dependence of ultrafast photoinduced reflectivity transients
is reported in Nd$_{0.5}$Sr$_{0.5}$MnO$_{3}$ thin film. The photoinduced
reflectivity shows a complex response with very different temperature
dependences on different timescales. The response on the sub-ps timescale
appears to be only weakly sensitive to the 270K-metal-insulator phase
transition. Below $\sim 160$ K the sub-ps response displays a two component
behavior indicating inhomogeneity of the film resulting from the substrate
induced strain. On the other hand, the slower response on the 10-100 ps
timescale is sensitive only to the metal-insulator phase transition and is
in agreement with some previously published results. The difference in the
temperature dependences of the responses on nanosecond and $\mu $s
timescales indicates that thermal equilibrium between the different degrees
of fredom is established relatively slowly - on a nanosecond timescale.
\end{abstract}

\section{Introduction}

The magnetoresistive (MR) manganites\cite
{SearleWang69,KustersSingelton89,HelmoltWecker93} of the type (Re,A)MnO$_{3}$
(Re and A are trivalent rare-earth and divalent ions respectively) have been
studied with a broad spectrum of experimental techniques including the
ultrafast pump-probe time-resolved spectroscopy.\cite
{MatsudaMachida98,LobadTaylor00,MerteljBosak00,FeibigMiyano00,RenZhang01,LobadAveritt01,AverittLobad01,OgasawaraKimura01,OgasawaraTobe02}
Most of the time-resolved studies were focused on the
metallic/magnetoresistive part of the phase diagram.\cite
{MatsudaMachida98,LobadTaylor00,MerteljBosak00,RenZhang01,LobadAveritt01,AverittLobad01}
while the charge ordered part of the phase diagram has been studied less
extensively\cite{FeibigMiyano00,OgasawaraKimura01,OgasawaraTobe02}.

Near mid-doping, where the ratio between number of Mn$^{3+}$ and Mn$^{4+}$
ions is around 1, there is a strong competition between the charge ordered
(CO) antiferromagnetic (AF) state and the metallic ferromagnetic (FM) state.
It has been shown that the CO state near mid-doping melts under intense
photoexcitation\cite{OgasawaraKimura01,OgasawaraTobe02}, while the low
photoexcitation dynamics has not been systematically studied in this doping
range.

In Nd$_{1-x}$Sr$_{x}$MnO$_{3}$ near mid-doping ($x\approx 0.5)$ there is a
very narrow region of $x$ where two successive phase transitions are
observed when the temperature is decreased below room temperature. The first
is the transition to the metallic state which is followed by the transition
to the CO phase at lower temperature.\cite{KajimotoYoshizawa99} The
stability of the CO phase is very sensitive to external perturbations such
as magnetic field\cite{KuwaharaTomioka95}, substrate-induced strain\cite
{PrellierBiswas99}, photoexcitation\cite{OgasawaraTobe02,Mori98}, etc.

To improve our understanding of the dynamics in the MR manganites and the
stability of the CO phase near mid-doping we applied the
low-excitation-fluence ultrafast pump-probe time-resolved spectroscopy to a
thin film of pseudocubic Nd$_{1-x}$Sr$_{x}$MnO$_{3}$ with $x\approx 0.5$, in
which CO is suppressed due to the substrate-induced strain. We observe a
different response on different timescales. The response on the 10-100-ps
timescale is similar to the response observed in the
metallic/magnetoresistive part of the phase diagram.\cite{LobadTaylor00} The
sub-ps response displays a two component behavior at lower temperatures
which we attribute to the layered inhomogeneity of the film due to the
substrate-induced strain in the film.

\section{Experimental}

A sample of Nd$_{1-x}$Sr$_{x}$MnO$_{3}$ thin film with nominal composition $%
x=0.5$ was deposited on an 110-oriented SrTiO$_{3}$ substrate by the aerosol
MOCVD\cite{GorbenkoKaul97}. The pseudocubic 110 direction of the thin film
was found to be parallel to the 110 direction of the substrate. The
temperature dependence of the magnetic moment parallel to the film and the
film resistance $R\,$, shown in Fig. 1, indicate a metal-insulator
transition (MIT) with an onset at $T_{MIT}\approx 265$ K and a peak of $%
dR/dT $ at $253$ K. Despite the nominal Sr concentration $x=0.5$ the
transition to the charge ordered (CO) state which is observed below $\sim
150 $ K in bulk samples was not observed in the resistivity measurements.
The absence of the CO transition could be either due to a minor discrepancy
of the real cation composition from the nominal one or more likely due to
the strain resulting from the film-substrate lattice constant mismatch. In
magnetic moment temperature dependence we observe a kink at 103K The kink is
absent in high magnetic field (45000 Oe) indicating that it is not due to
the background susceptibility of the SrTiO$_{3}$ substrate which has a
structural phase transition in the range 80-105K depending on impurities.

We used a mode locked Ti:sapphire laser operating at $\lambda $ = 800 nm
(1.5 eV) as a source of pump and probe pulses. The train of $\sim $%
60-fs-long pump laser pulses was focused on a 60-$\mu $m-diameter spot
giving an optical fluence of $\sim $5 $\mu $J/cm$^{2}$ per pulse. The train
of appropriately delayed probe pulses, which had the fluence 50-100 times
weaker than the pump pulses, was reflected from the same spot and detected
by a photodiode. The probe pulses were polarized orthogonally to the pump
polarization and a polarization analyzer which was oriented perpendicularly
to the pump polarization was mounted in front of the photodiode to eliminate
the signal due to pump scattering. High frequency modulation (200 kHz) of
the pump beam intensity enabled low-noise lock-in detection of the
photoinduced reflectivity $\Delta \Re /\Re $ (PIR). The thickness of the
film was approximately one optical penetration depth estimated from its
optical density. Other details of the experimental pump-probe setup were
published elsewhere.\cite{MihailovicDemsar99}

The sample was mounted on a cold finger of an optical liquid-He flow
cryostat. Due to laser heating, the temperature of the illuminated sample
volume was estimated to be less than 15 K above the cryostat temperature\cite
{Demsar99}, which was directly measured using a RhFe resistor.

We observed no change of PIR transients when the pump polarization is
rotated relative to the thin-film crystal-axes. A typical measured PIR
transient is shown in Fig. 2. The transient is characterized by three delay
regions annotated by capital letters A,B and C in Fig. 2. At negative delay
times (region A) we observe a delay-independent PIR background, which
represents the PIR build-up due to relaxation/diffusion processes
significantly longer than the inter-pulse separation of 12.5 ns. The
PIR-background amplitude is temperature dependent as shown in Fig. 3. The
lock-in out-of-phase signal, which is delay-independent within the full
delay range, is also shown in the same plot for comparison. The out-of-phase
lock-in signal actually corresponds to the PIR which persists during the
''dark'' part of the 200-kHz pump modulation period. It has virtually the
same temperature dependence as the PIR background, with the only difference
being a different sign, implying a temperature independent $-\pi /4$ phase
shift of the induced reflectivity at 200-kHz with respect to the pump
modulation. The timescale associated with the PIR background is therefore of
the order of $\mu $s$.$ The magnitudes of both signals exhibit a peak at $%
\sim $240 K and the signals change sign at $\sim $260 K. Taking into account
the $\sim $10-K temperature difference of the illuminated sample volume
relative to the cryostat temperature the background-peak temperature roughly
coincides with the $dR/dT$-peak while the sign change coincides with $%
T_{MIT} $.

To aid the analysis of the PIR\ transients on the sub-ns timescale we
substract the corresponding delay independent background offsets discussed
above. We refer to the modified transients as ultrafast-PIR (UPIR)
transients. In Fig. 4 we show UPIR transients at different temperatures. We
observe two well separated timescales (region B and region C in Fig. 2 and
Fig. 4) in which the transients show different temperature dependences. The
regions are separated by a break in Fig. 4. Region B includes dynamics up to 
$\sim 1$ ps while region C includes slower dynamics up to $\sim $250 ps.

Let us first look closer at the temperature dependence of the slower
transients (region C). At room temperature, the UPIR transient is negative
and virtually flat for delays larger than $\sim $1 ps. When the temperature
is decreased below 260 K a new exponential relaxation component appears,
leading to a change of sign of the UPIR transient at longer delay times. The
temperature dependences of the relaxation time and the amplitude of the
component are plotted in Fig. 5.

At maximum delay (250 ps) the UPIR transient magnitude is finite in the
whole temperature range except around $\sim 260$ K indicating the presence
of a longer, nanosecond-timescale relaxation process. Similar to the PIR
background, it exhibits a change of sign at $\sim $260 K but is virtually
temperature independent below 240 K as shown in Fig. 6.

Let us now discuss the ultrafast response {\em above }$160${\em \ K} for
delays up to $\sim 1$ ps. In this range the UPIR transient is negative ($%
\Delta R/R<0$) with a 130-fs-long leading edge which is followed by a $\sim
300$ fs exponential decay continuing into the slow relaxation component
described above. The $130$-fs risetime is temperature independent in this
range. The magnitude of the UPIR transient peak located at $\sim 100$ fs
increases with decreasing temperature exhibiting a broad extremum at 240 K
and decreasing again when the temperature is further decreased as seen in
Fig. 7 (b). The $\sim 300$-fs relaxation time is only weakly sensitive to
the MIT as seen from Fig. 7 (a) showing a slight increase when the
temperature is lowered towards $T_{MIT}$ and a slight decrease below $%
T_{MIT} $. Unfortunately, additional noise due to the large PIR background,
which appears in this temperature range, \thinspace causes scatter in the
data and prevents a more precise determination of the fast relaxation-time
temperature dependence.

When the temperature is decreased {\em below }$\sim 160${\em \ K} the sub-ps
component experiences a remarkable change in shape (see Fig. 4). Firstly
between 160-100 K the leading edge shifts together with the peak by $\sim
100 $ fs towards the positive delay. In addition, the peak broadens and at
80 K disappears. Below 80 K a positive and slightly steeper leading edge
centered again at the zero-delay is observed followed by a positive peak at $%
\sim 50$ fs delay and a minimum around $\sim 200$ fs. After the minimum the
UPIR transients still increase slightly on a timescale of $\sim 300$ fs
followed by the slow relaxation component already described. The major
changes of the shape of the sub-ps UPIR transients appear at similar
temperature as the kink in the magnetic moment temperature dependence when
the $\sim $10-K laser heating of the illuminated volume is taken into
account.

\section{Discussion}

The 1.5 eV photon energy used in the present experiment lies in the region
of optical transitions between split Mn $e_{g}\,$derived bands\cite
{QuijadaCerne98}. The time resolved optical reflectivity at 1.5 eV photon
energy is therefore directly related to the dynamics of charge carriers,
which are the most relevant for the physics of MR manganites. In addition,
in Nd$_{1-x}$Sr$_{x}$MnO$_{3}$ the reflectivity edge due to the coherent
response at low temperatures is below the probe photon energy. A
contribution to the photoinduced reflectivity from the collective charge
response due to the free carriers at low temperatures is therefore expected
not to be significant in comparison to the interband contribution.

Let us now discuss the different delay regions separately, proceeding from
the fastest to the slowest timescale.

{\em Above} $\sim 160$ K the 300-fs relaxation time in our sample is just
slightly longer than the 200-fs relaxation time measured by us in La$_{0.82}$%
Pb$_{0.18}$MnO$_{3}$\cite{MerteljBosak00} and La$_{1-x}$Sr$_{x}$MnO$_{3}$%
\cite{Mertelj}. In La$_{1-x}$Ca$_{x}$MnO$_{3}$ Lobad {\em et al.}\cite
{LobadTaylor00} report a sub-100-fs absorption transient, but do not discus
the transient shape in detail. Since our time resolution is $\sim 80$ fs the
200-300-fs relaxation times measured in\ Nd$_{0.5}$Sr$_{0.5}$MnO$_{3}$ and La%
$_{0.82}$Pb$_{0.18}$MnO$_{3}$ are intrinsic to the samples. However, any
additional sub-80-fs components are not completely excluded by our
measurements and it is not clear whether significant differences in {\em the
relaxation time} exists between different manganites.

In contrast, {\em the amplitude} of the sub-ps relaxation component clearly
behaves differently below $T_{MIT}$ in manganites with different
compositions on the La site. In La$_{0.82}$Pb$_{0.18}$MnO$_{3}$\cite
{MerteljBosak00} the amplitude increases with decreasing temperature, in La$%
_{0.7}$Sr$_{0.3}$MnO$_{3}$\cite{Mertelj} the increase in amplitude is less
pronounced while in Nd$_{0.5}$Sr$_{0.5}$MnO$_{3}$ the amplitude even
decreases below $T_{MIT}$. It remains to be more systematically investigated
whether this is correlated with the hole doping, as current data suggest, or
it is simply connected with the cation species type and/or the ionic radius
on the La site.

{\em Below} $\sim 160$ K the sub-ps behavior in our Nd$_{0.5}$Sr$_{0.5}$MnO$%
_{3}$ thin film is much more complex than in manganite thin films with
smaller hole doping.\cite{MerteljBosak00,Mertelj} Despite anomalies observed
in the ultrafast optical response and the magnetic moment temperature
dependence there is no signature of any anomaly in the resistivity
temperature dependence. This can be explained by the layered inhomogeneity
of the film due to the substrate-film lattice-mismatch strain. It has been
shown that thin films of Nd$_{0.5}$Sr$_{0.5}$MnO$_{3}$ deposited on LaAlO$%
_{3}$ substrate display a layered structure with a $\sim 20$ nm thick
insulating layer near the substrate and a less strained layer on top, which
undergoes MIT.\cite{PrellierBiswas99} In addition the remaining strain in
the top layer of the Nd$_{0.5}$Sr$_{0.5}$MnO$_{3}$ film is sufficient to
almost completely suppress the CO transition.\cite{PrellierBiswas99} Similar
behavior was observed also in a La$_{0.66}$Sr$_{0.33}$MnO$_{3}$ thin films.%
\cite{BiswasRajeswari01} In our case the substrate lattice parameter is $%
\sim 1.5\%$ larger than the average pseudocubic lattice parameter of bulk Nd$%
_{0.5}$Sr$_{0.5}$MnO$_{3}$. Despite the strain in our case has different
sign than in the case of LaAlO$_{3}$ substrate, the absence of any anomaly
in the resistance temperature dependence suggests a similar layered
inhomogeneity of the film. Since the thickness of the film is of the order
of the optical penetration depth optical (and magnetization) measurements
detect contributions from top and bottom layers while in the electrical
transport measurement the top high-conductivity metallic layer effectively
shortens out the bottom insulating layer below MIT.

We therefore attribute anomalies observed below $\sim 160$ K in the sub-ps
UPIR transients to the bottom insulating layer(s). We analyze the transients
below $\sim 160$ K in terms of an additional positive sub-ps
exponentially-relaxing component originating from the bottom insulating
layer. The component partially cancels the leading edge of the negative
300-fs component observed above $\sim 160$ K reproducing the observed
leading edge shift, as can be seen from the solid lines in Fig. 4.
Unfortunately it is not possible to extract neither amplitudes nor
relaxation-times temperature dependences of both components reliably since
the fits are extremely ill-conditioned. We can therefore only estimate the
relaxation time of the additional component to be below 100 fs while the
relaxation time of the negative one remains in the 200-300 fs range.

Let us now discuss the origin of the anomaly attributed to the bottom
insulating layer. The SrTiO$_{3}$ substrate undergoes a second order
structural phase transition from cubic to tetragonal symmetry at $\sim 105\,$%
K. The temperature range at which major anomalies are observed in our Nd$%
_{0.5}$Sr$_{0.5}$MnO$_{3}$ thin film are very similar to this temperature.
However, the leading edge shift of the UPIR is already evident in the 130-K
UPIR transient well above the substrate structural phase transition. In
addition the substrate lattice parameters splitting below the structural
transition is less than $0.1\%$\cite{HirotaHill95} which is more than ten
times less then the substrate-film lattice mismatch. We therefore believe
that the observed anomaly is not purely substrate induced effect but that it
is related to the Nd$_{0.5}$Sr$_{0.5}$MnO$_{3}$ instability towards CO.
Unfortunately from our data alone we can not determine whether the
additional sub-100 fs relaxation is indeed related to CO since there is no
data yet available on time resolved PIR at low excitation densities on
charge ordered manganites.

Proceeding towards longer timescales we look next at region B. The
relaxation on a 100-ps timescale observed in La$_{1-x}$Ca$_{x}$MnO$_{3}$ and
La$_{1-x}$Sr$_{x}$MnO$_{3}$ was attributed to the magnetization relaxation
dynamics by Lobad {\em et al.} \cite{LobadTaylor00}, while Ren {\em et al.}%
\cite{RenZhang01} attribute it to polaron relaxation dynamics. It still
needs to be confirmed - for example by time resolved Faraday rotation
measurements - whether the hundred-ps component is indeed related directly
to the magnetization relaxation as proposed by Lobad {\em et al.}\cite
{LobadTaylor00}. In case of Nd$_{0.5}$Sr$_{0.5}$MnO$_{3}$ we observe a
similar temperature dependence of the relaxation time of the hundred-ps
component as Lobad {\em et al.}\cite{LobadTaylor00} (Fig. 5), although the
relaxation is somewhat faster in Nd$_{0.5}$Sr$_{0.5}$MnO$_{3}$.

For comparison, in (La,Pb)MnO$_{3}$\cite{MerteljBosak00} the peaks below $%
T_{MIT}$ in the relaxation-time and amplitude temperature dependences of the
hundred-ps component are not observed. Instead, the relaxation time and the
amplitude below $T_{MIT}$ steadily increase with the decreasing temperature
with a trace of a hump more than 100K below $T_{MIT}$. In La$_{1-x}$Ca$_{x}$%
MnO$_{3}$ Ren {\em et al.}\cite{RenZhang01} report a monotonic decrease of
the relaxation time below $T_{MIT}$. While in La$_{1-x}$Ca$_{x}$MnO$_{3}$
the different behavior may be related to the higher fluence used in their
experiment, it is not clear at the moment what is the origin of the observed
difference in (La,Pb)MnO$_{3}$. One possible reason for the absence of the
peak below $T_{MIT}$ in (La,Pb)MnO$_{3}$ is a surplus oxygen content of the
film leading to cation vacancies on the Mn and La sites. The disorder could
cause faster relaxation due to a broken translation symmetry, but it could
also lead to low-lying long-lived localized states. The contribution from
such states can not be excluded on the basis of current data on (La,Pb)MnO$%
_{3}$.\cite{MerteljBosak00}

The residual UPIR transient amplitude at 250 ps measures the amount of decay
between successive laser pulses (12.5 ns). Its temperature dependence is
similar to that of the PIR background on the $\mu $s timescale with the sign
change at 260 K, but without the peak below $T_{MIT}$. We observe a similar
behavior in (La,Pb)MnO$_{3}$.\cite{MerteljBosak00} The absence of the peak
below $T_{MIT}$ indicates that the response on the nanosecond timescale is
(at least in part) {\em not bolometric} in origin, while the change of sign
at 260K suggests that the optical transition contributing to PIR on both
timescales is the same.

Let us finally discuss the $\mu $s timescale corresponding to the large
temperature dependent PIR background. The slowly decaying signal observed at
negative delay times (region A) seems to be a general feature present in
transient PIR on the GMR manganites observed by us\cite{MerteljBosak00} and
others\cite{LobadTaylor00}. Its temperature dependence exhibits a peak at
the temperature, which is usually of the order of 10 K below $T_{MIT}$, with
the exception of (La,Pb)MnO$_{3}$, where the peak is $\sim 100$ K below $%
T_{MIT}$.\cite{MerteljBosak00}

A similar background, albeit with a smaller magnitude, was observed in the
high $T_{c}$\thinspace cuprates\cite{StevensSmith97} and the charge density
wave (CDW) semiconductor K$_{0.3}$MoO$_{3}$ \cite{DemsarBiljakovic99}. In
these cases the background is peaked at the superconducting/CDW $T_{c}$ and
was attributed to the intragap localized states\cite{KabanovDemsar00}. No
corresponding out-of-phase lock-in signal similar to that in the manganites
was observed in either the cuprates or K$_{0.3}$MoO$_{3}$. Despite the fact
that the behavior of the background in manganites is somewhat different, we
have previously tentatively attributed the PIR background to a similar
mechanism.\cite{MerteljBosak00} A detailed analysis of the data presented in
this paper suggests that this assignment for {\em manganites} needs to be
reconsidered, since in manganites the PIR-background temperature dependence
and the constant $-\pi /4$ background phase shift are more consistent with a
bolometric response as explained below.

The expected amplitude of the bolometrically induced reflectivity change is
given by 
\begin{equation}
\frac{\Delta \Re }{\Re }=\frac{1}{\Re }\frac{\partial \Re }{\partial T}%
\Delta T,  \label{bolometric}
\end{equation}
where $\Delta T$ is the temperature oscillation amplitude. The specific heat
and the thermal conductivity of the substrate and the manganite film\cite
{ParkJeong97,CohnNeumeier97} below MIT are both relatively smooth functions
of temperature without any significant anomalies. $\Delta T$ is therefore
expected to be a relatively smooth function of temperature, and can
contribute neither to the change of the PIR-background sign at 260K nor to
the observed peak below $T_{MIT}$. On the other hand, the reflectivity
around 1.5 eV in manganites is virtually constant above $T_{MIT}$, but
exhibits a systematic drop when the temperature is lowered below $T_{MIT}$
with further, albeit slower, decrease towards lower temperatures\cite
{SaitohOkimoto99,LeeJung00}. The shape of the PIR background temperature
dependence in Fig. 3 is therefore consistent with the temperature dependence
of the derivative $\frac{1}{\Re }\frac{\partial \Re }{\partial T}$ which has
a peak just below $T_{MIT}$ and is very close to zero or my be even negative
above $T_{MIT}$.

To estimate the maximum magnitude of the bolometrically induced reflectivity
change we take the available data for Nd$_{0.7}$Sr$_{0.3}$MnO$_{3}$\cite
{LeeJung00} where the reflectivity drop below $T_{MIT}$ is $\sim 0.04$ in
the temperature range of $\sim 30K$. A similar reflectivity drop is expected
in Nd$_{0.5}$Sr$_{0.5}$MnO$_{3}$ judging from a similar temperature
dependence of the optical conductivity.\cite{JungLee00} Taking into account
the estimated 0.5K temperature oscillation amplitude\cite{deltat} we obtain $%
\frac{\Delta \Re }{\Re }\approx 4\cdot 10^{-3}$ for the maximum in-phase
bolometrically induced reflectivity change. This is in a fair agreement with
the observed maximum amplitude of the PIR background which is $\sim 10^{-3}$%
. The PIR background amplitude and its temperature dependence are therefore
consistent with a bolometric response.

Further, the temperature independent phase shift of $-\pi /4$ between
harmonically modulated average laser intensity and the surface temperature
oscillation observed at 200 kHz in our experiment is a characteristic of a
1D diffusion process.\cite{CarslawJaeger59} In our case the heat diffusion
length on a timescale of the lock-in modulation frequency is a few $\mu $m
as estimated from typical heat capacity and thermal conductivity data for
manganites\cite{CohnNeumeier97,ParkJeong97} and the substrate. Since the
laser beam diameter of 60$\mu $m is much larger than the heat diffusion
length at the modulating frequency the heat diffusion is indeed expected to
be effectively 1D, further supporting the bolometric origin of the
background PIR in the manganites.

A similar $-\pi /4$ temperature independent phase shift observed in
(La,Pb)MnO$_{3}$\cite{MerteljBosak00} suggests the bolometric origin of the
PIR background in that case as well.

\section{Conclusion}

We measured temperature dependence of ultrafast photoinduced reflectivity
transients in Nd$_{0.5}$Sr$_{0.5}$MnO$_{3}$ thin film. On the sub-ps
timescale the transients are only weakly sensitive to the metal-insulator
transition at $\sim $\thinspace $270$ K, but instead show a remarkable
change in shape below $\sim 160$ K. Major changes in the shape are
correlated with an anomaly observed in the magnetization temperature
dependence, but no anomaly is observed in the temperature dependence of the
resistivity in this temperature range. The absence of the anomaly in
resistivity is attributed to the substrate-film lattice-mismatch induced
inhomogeneity of the film. The changes of the transients could be attributed
to an additional sub-100-fs relaxation component, which originates from the
most strained layer of the film near the substrate. Although the substrate
undergoes a structural phase transition in the same temperature range, the
small changes of the substrate lattice parameters below the phase transition
and the higher onset-temperature of the changes in the transients suggest
that the observed anomalies are not due to the substrate but are related to
the CO instability of bulk Nd$_{0.5}$Sr$_{0.5}$MnO$_{3}$.

The slower 10-100-ps photoinduced-reflectivity relaxation dynamics is
sensitive only to the $\sim $\thinspace $270$-K MIT exhibiting peaks in
relaxation time and amplitude just below the MIT temperature, similar as in
some previously studied manganites with different doping. \cite
{LobadTaylor00}

On even slower timescales the photoinduced reflectivity is also sensitive to
the $\sim $\thinspace $270$-K MIT only. While the ultra-slow-response
temperature dependence on the $\mu $s timescale is consistent with the
expected bolometric response, the relaxation on the 1-10-ns timescale shows
a different temperature dependence and is most probably not bolometric. This
indicates that the thermal equilibrium between different degrees of freedom
(charge, spin) is not reached faster than on the nanosecond timescale.

\section{Figure Captions}

Figure 1. The film magnetic moment and resistance as functions of the
temperature. The magnetic moment was measured in the external magnetic field
of 10 Oe parallel to the film.

Figure 2. A typical measured PIR transient (a). The transient is
characterized by three delay regions annotated by capital letters A,B and C
(b). Note logarithmic scale after the break.

Figure 3. The photoinduced reflectivity magnitude at negative delay with the
lock-in out-of-phase signal shown for comparison.

Figure 4. Ultrafast photoinduced reflectivity transients at different
temperatures. Dotted and continuous lines represent measured data and fits
respectively. The delay independent PIR\ background was substracted and
traces were shifted vertically for clarity. Note the logarithmic timescale
after the break.

Additional noise present in transients below 250 K is a consequence of the
substracted delay independent PIR\ background increase below $T_{MIT}$.
Around $\sim $240 K its magnitude is almost ten times larger than the UPIR\
transient magnitude.

Figure 5. The relaxation time (a) and the magnitude (b) of the slow
relaxation component (region C) as functions of the temperature. Error bars
indicate standard error estimated from fits.

Figure 6. The magnitude of the UPIR transients at the 250 ps delay as a
function of temperature. Error bars were estimated from noise present in the
transients.

Figure 7. The relaxation time of the fast sub-ps decay (region B) (a) and
the sub-ps peak amplitude as functions of the temperature above 150K (b).
Error bars in (a) indicate standard error estimated from fits while in (b)
error bars were estimated from noise present in the transients.


\begin{references}
\bibitem{SearleWang69}  C.W. Searle, S.T. Wang, {\em Can. J. Phys.} {\bf 47}%
, 2703 (1969).

\bibitem{KustersSingelton89}  R.M. Kusters, J. Singelton, D.A. Ken, R.
McGreevy, W. Hayes, {\em Physica} {\bf B 155}, 362 (1989).

\bibitem{HelmoltWecker93}  R. von Helmolt, J. Wecker, B. Holzapfel, L.
Schultz, K. Samwer, {\em Phys. Rev. Lett.} {\bf 71}, 2331 (1993).

\bibitem{MatsudaMachida98}  K. Matsuda, A. Machida, Y. Moritomo, A.Nakamura, 
{\em Phys. Rev.} {\bf B 58}, 4203 (1998).

\bibitem{LobadTaylor00}  A.I. Lobad, A.J. Taylor, C. Kwon, S.A. Trugman, T.
R. Gosnell, {\em Chem. Phys. }{\bf 251}, 227 (2000).

\bibitem{MerteljBosak00}  T. Mertelj, A.A, Bosak, O. Yu. Gorbenko, A.R.
Kaul, D. Mihailovic, {\em Int. J. Mod. Phys. B }{\bf 14}, 3584 (2000).

\bibitem{FeibigMiyano00}  M. Feibig, K. Miyano, Y. Tamioka, Y. Tokura, {\em %
Appl. Phys. B} {\bf 71}, 211 (2000).

\bibitem{RenZhang01}  Y.H. Ren, X. H. Zhang, G. L\"{u}pke, M. Schneider, M.
Onellion, I.E. Perakis, Y.F. Hu, Qi Li, {\em Phys. Rev.} {\bf B 64}, 144401
(2001).

\bibitem{LobadAveritt01}  A.I. Lobad, R.D. Averitt, A.J. Taylor, {\em Phys.
Rev.} {\bf B 63}, 060410 (2001).

\bibitem{AverittLobad01}  R.D.Averitt, A.I. Lobad, C. Kwon, S.A. Trugman,
V.K. Thorsm\o lle, A.J. Taylor, {\em Phys. Rev. Lett.} {\bf 87}, 017401
(2001).

\bibitem{OgasawaraKimura01}  T. Ogasawara, T. Kimura, T. Ishikawa, M.
Kuwata-Gonokami, Y. Tokura, {\em Phys. Rev.} {\bf B 63}, 113105 (2001).

\bibitem{OgasawaraTobe02}  T. Ogasawara, K. Tobe, T. Kimura, H. Okamoto, Y.
Tokura, J. Phys. Soc. Jpn {\bf 71}, 2380 (2002).

\bibitem{KajimotoYoshizawa99}  R. Kajimoto, H. Yoshizawa, H. Kawano, H.
Kuwahara, Y. Tokura, K. Ohoyama, and M. Ohashi, {\em Phys. Rev.} {\bf B 60},
9506 (1999).

\bibitem{KuwaharaTomioka95}  H. Kuwahara, Y. Tomioka, A. Asamitsu, Y.
Moritomo, and Y. Tokura, {\em Science} {\bf 270}, 961 (1995).

\bibitem{PrellierBiswas99}  W. Prellier, A. Biswas, M. Rajeswari, T.
Venkatesan, R.L. Greene, {\em Appl. Phys. Lett. }{\bf 75}, 397 (1999).

\bibitem{Mori98}  T. Mori, Phys. Rev. B 58, 12 543 (1998).

\bibitem{GorbenkoKaul97}  O. Yu. Gorbenko, A. R. Kaul, N. A. Babuskina, L.
M. Belova,{\em \ J. Mater. Chem }{\bf 7}, 747 (1997).

\bibitem{MihailovicDemsar99}  D. Mihailovic, J. Demsar, {\em Time-Resolved
Optical Studies of Quasiparticle Dynamics} . In {\em Spectroscopy of
Superconducting Materials} ed. by Eric Faulques (American Chemical Society,
Washington, DC, 1999).

\bibitem{Demsar99}  J. Demsar, {\em Photoexcited Carrier Relaxation in HTC
Superconductors probed by Ultrafast Optical Spectroscopy}, PhD. thesis
(Faculty of Math. and Phys., University of Ljubljana, 1999).

\bibitem{BiswasRajeswari01}  A. Biswas, M. Rajeswari, R.C. Srivastava, T.
Venkatesan, R.L. Greene, Q. Lu, A.L. deLozanne, A.J. Millis, {\em Phys. Rev. 
}{\bf B 63}, 184424 (2001).

\bibitem{HirotaHill95}  K. Hirota, J.P. Hill, S.M. Shapiro, G. Shirane, Y.
Fujii, {\em Phys. Rev. }{\bf B 52}, 13195 (1995).

\bibitem{QuijadaCerne98}  M. Quijada, J.Cerne, J.R. Simpson, H.D. Drew, K.H.
Ahn, A.J. Millis, R. Shreekala, R. Ramesh, M. Rajeswari, T. Venkatesan, {\em %
Phys. Rev. }{\bf B58} (1998) 16093.

\bibitem{StevensSmith97}  C. J. Stevens, D. Smith, C. Chen, and J. F. Ryan,
B. Podobnik and D. Mihailovic,\thinspace G. A. Wagner and J. E. Evetts, {\em %
Phys. Rev. Lett.}{\bf \ 78}, 2212 (1997).

\bibitem{DemsarBiljakovic99}  J. Demsar, K. Biljakovi\'{c}, D. Mihailovic, 
{\em Phys. Rev. Lett.}{\bf \ 83}, 800 (1999).

\bibitem{KabanovDemsar00}  V.V. Kabanov, J. Demsar, D. Mihailovic, {\em %
Phys. Rev. }{\bf B 61}, 1477 (2000).

\bibitem{ParkJeong97}  S. H. Park, Y.-H. Jeong, K.-B. Lee, S. J. Kwon, {\em %
Phys. Rev.} {\bf B 56}, 67 (1997).

\bibitem{CohnNeumeier97}  J. L. Cohn, J. J. Neumeier, C. P. Popoviciu, K. J.
McClellan, Th. Leventouri, {\em Phys. Rev.} {\bf B 56}, 8495 (1997).

\bibitem{SaitohOkimoto99}  E. Saitoh, Y. Okimoto, Y. Tomioka, T. Katsufuji,
Y. Tokura, {\em Phys. Rev.} {\bf B 60}, 10362 (1999).

\bibitem{LeeJung00}  H. J. Lee, J. H. Jung, Y. S. Lee, J. S. Ahn, and T. W.
Noh,\thinspace K. H. Kim, S-W. Cheong, {\em Phys. Rev.} {\bf B 60}, 5251
(2000).

\bibitem{JungLee00}  J. H. Jung, H. J. Lee, T. W. Noh, E. J. Choi, Y.
Moritomo, Y. J. Wang, X. Wei, {\em Phys. Rev.} {\bf B 62}, 481 (2000).

\bibitem{deltat}  Since the thicknes of the film is much smaller than the
heat diffusion length on the $\mu $s timescale, the heat diffussion out of
the experimental volume is governed by the heat diffusivity in the film and
in the substrate and the thermal resistance of the boundary beetwen the film
and the substrate. For purpose of estimation we neglect the boundary heat
resistivity, assume a negligible film thickness and calculate the surface
temperature oscilation amplitude and phase in a simple model assuming a
semi-infinite homogenous material using the substrate thermal constants and
the thin film optical constants.

\bibitem{CarslawJaeger59}  H.S. Carslaw, J.C. Jaeger, {\em Conduction of
Heat in Solids}, (Oxford University Press, Ely House, London, 1959).

\bibitem{Mertelj}  T. Mertelj, V.A. Dediu, unpublished data on La$_{0.7}$Sr$%
_{0.3}$MnO$_{3}$.
\end{references}
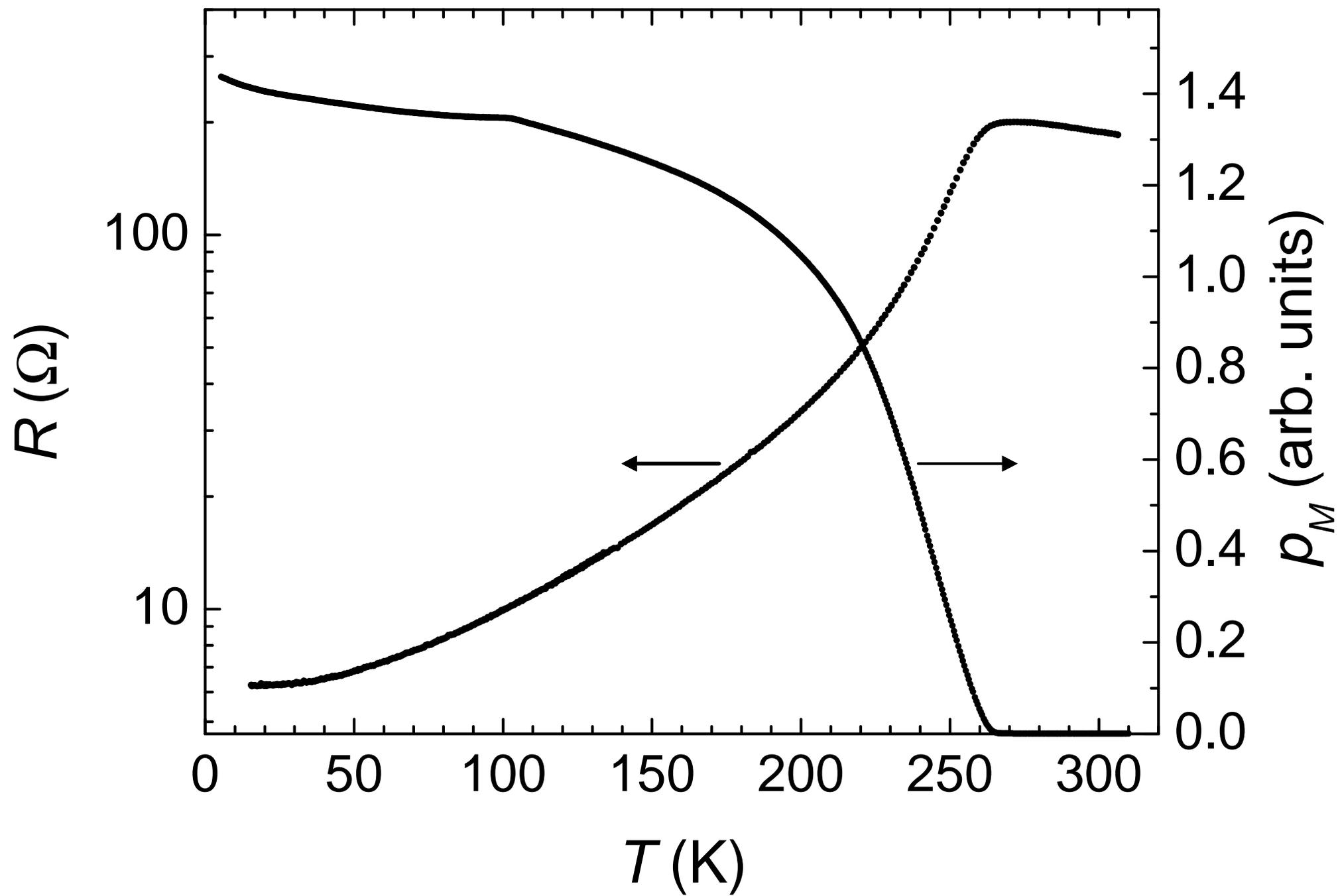
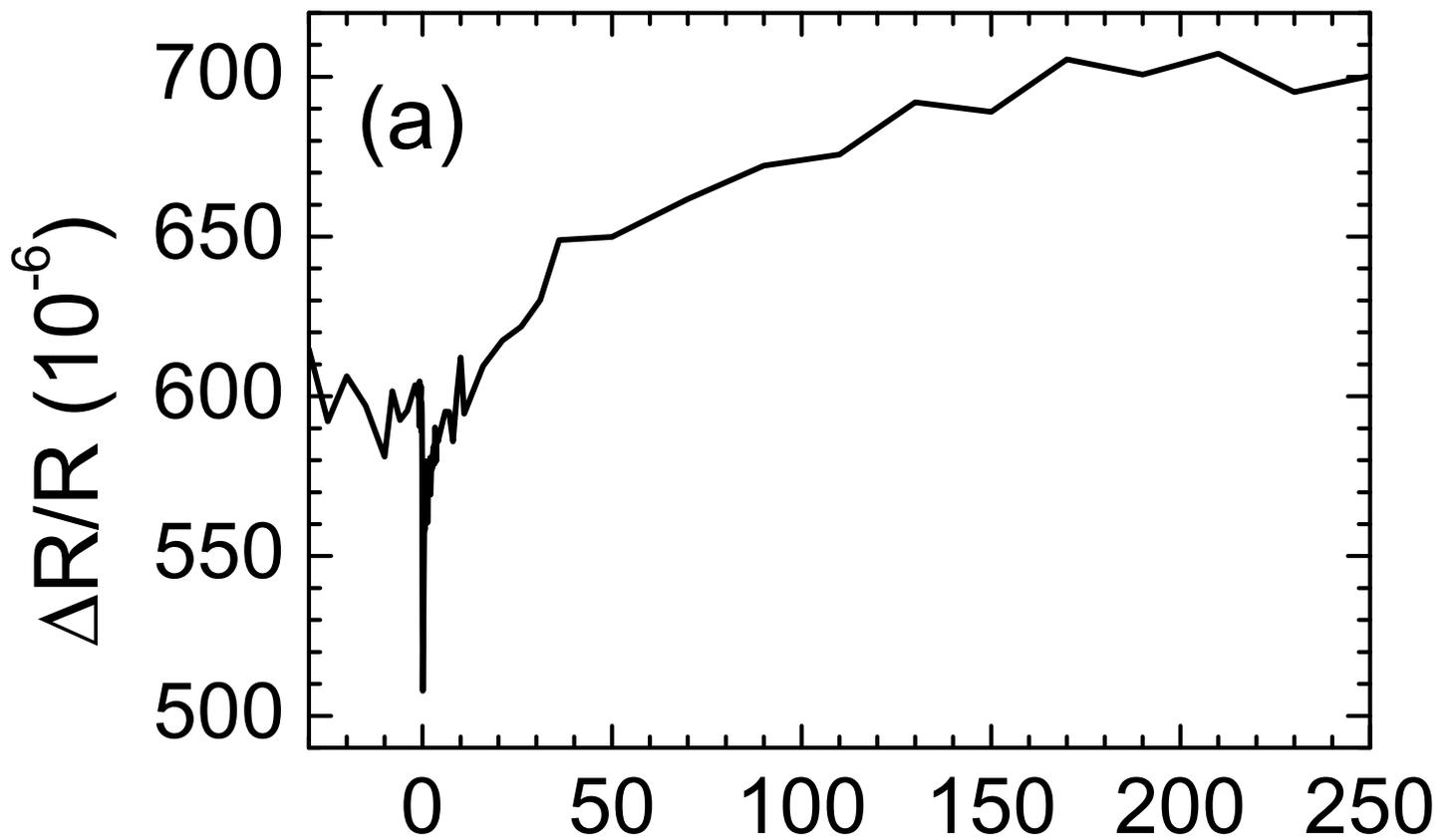
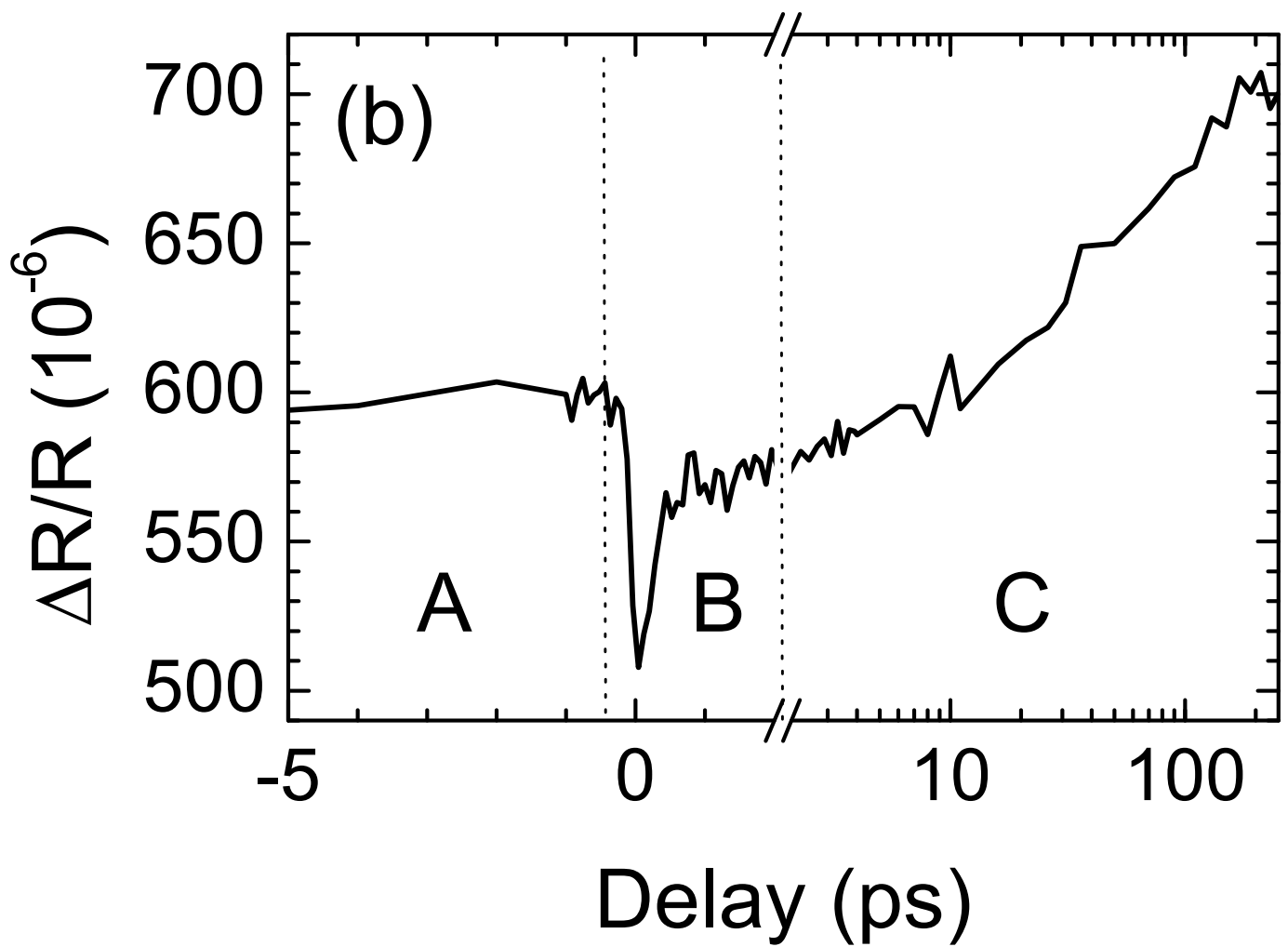
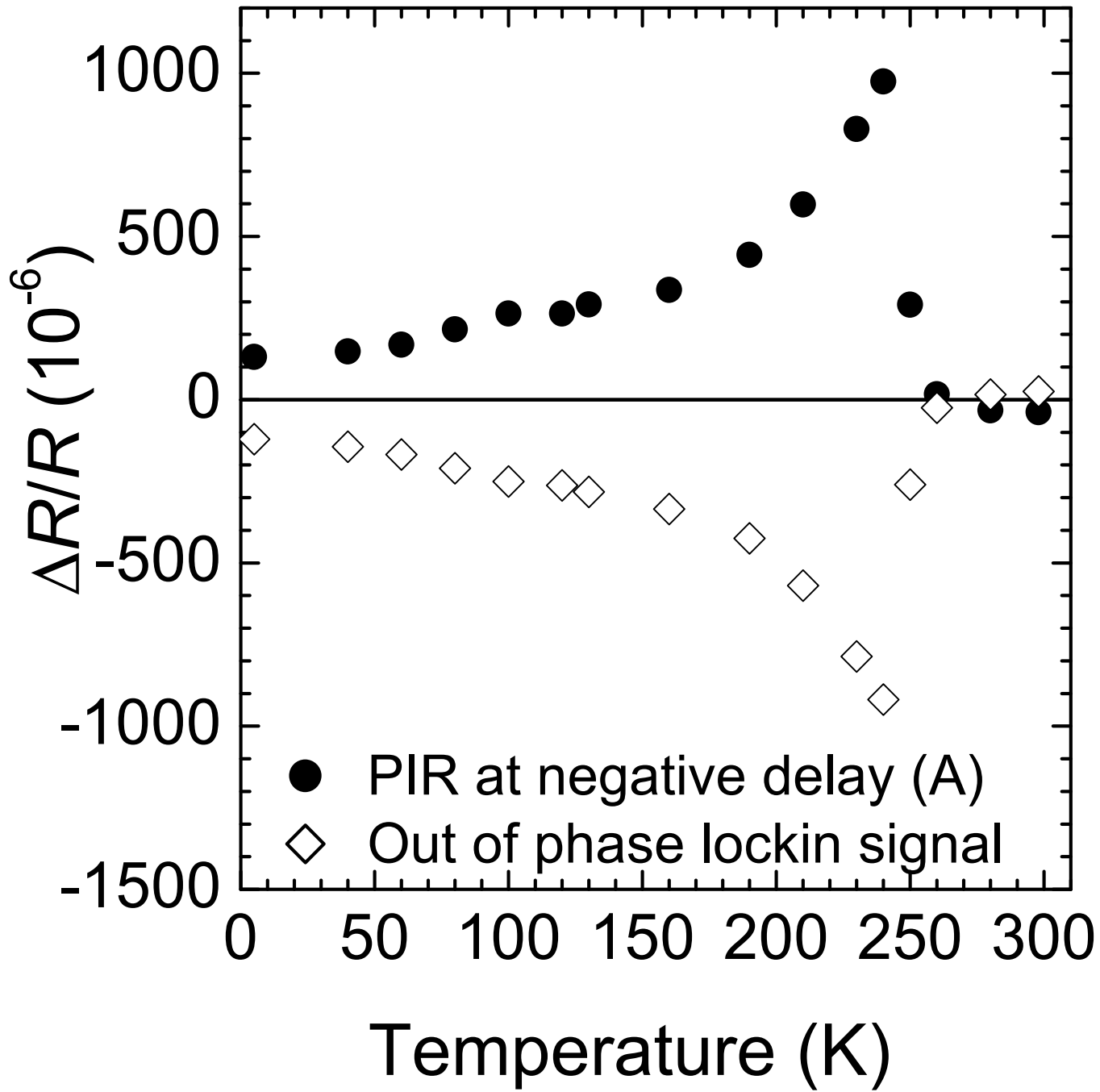
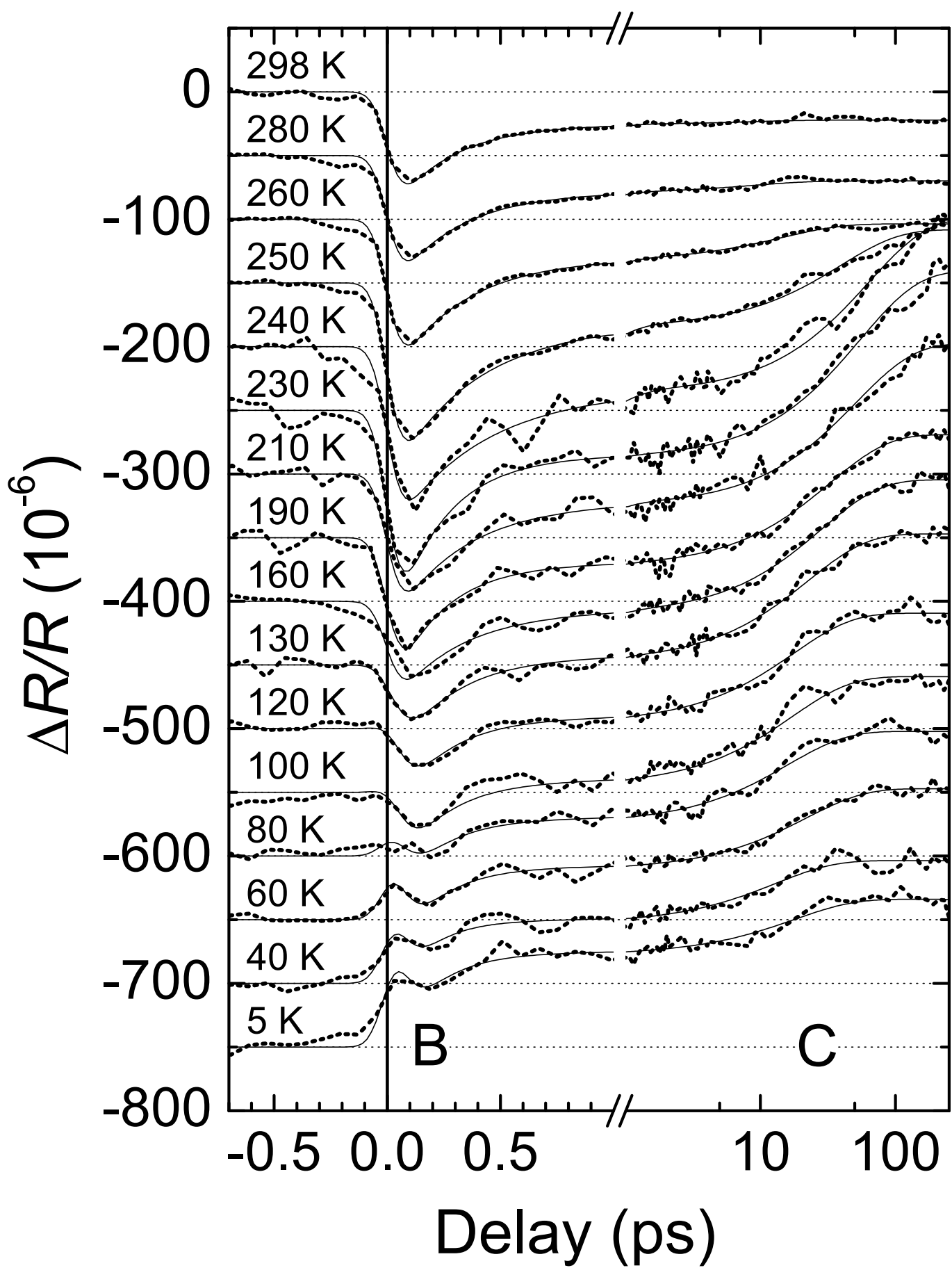
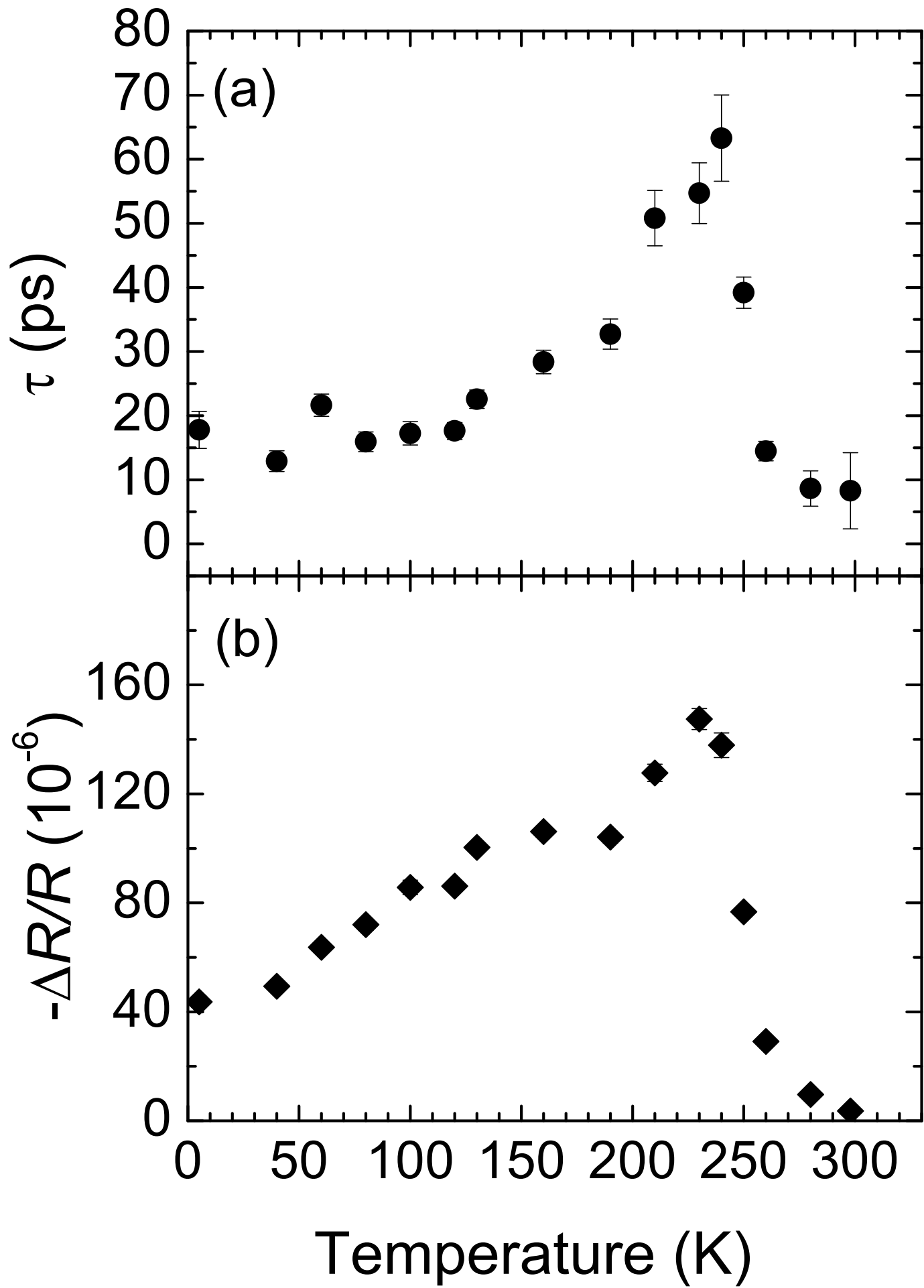
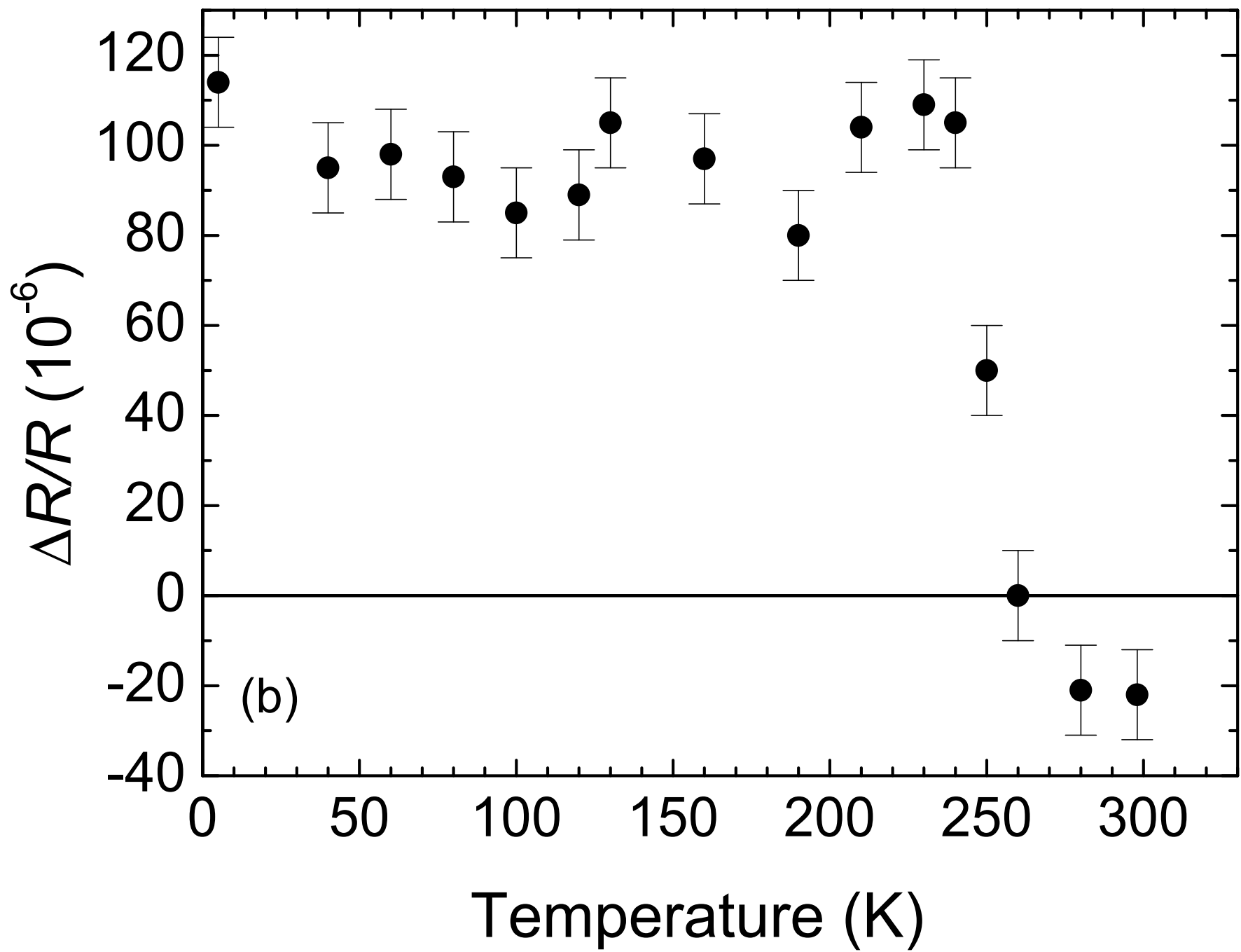
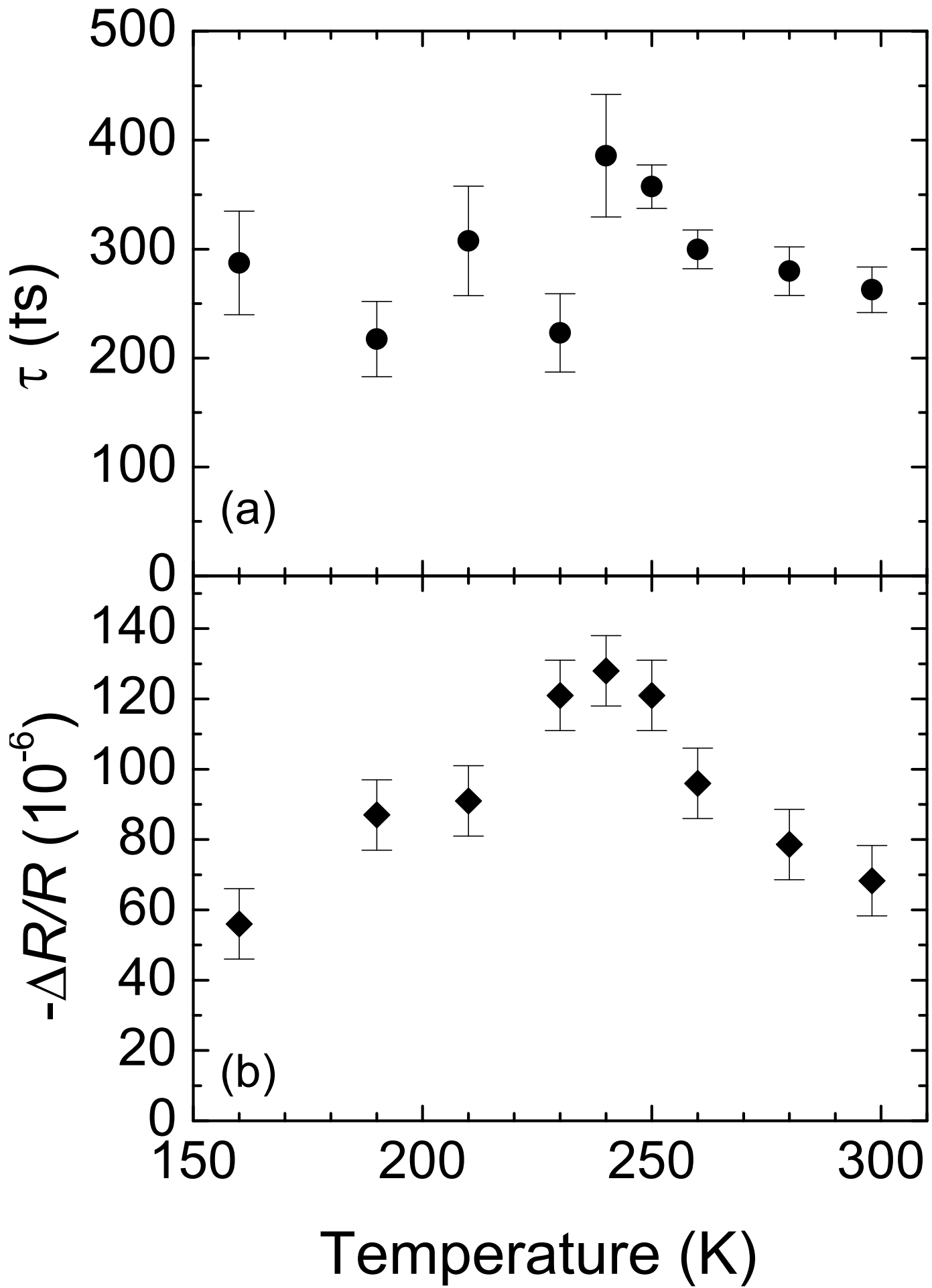
\end{document}